\documentclass[notitlepage, final, 10pt, aps, prl, floatfix, twocolumn, letter, english, superscriptaddress, longbibliography]{revtex4-2}
\pdfoutput=1

\usepackage[T1]{fontenc}
\usepackage[utf8]{inputenc}
\usepackage[english]{babel}
\usepackage{lmodern}
\usepackage{amsfonts}
\usepackage{amssymb}
\usepackage{amsmath}
\usepackage{bm}
\usepackage{bbm}
\usepackage{cases}
\usepackage{graphicx}
\usepackage{subfigure}
\usepackage{textgreek}
\usepackage{xcolor}
\definecolor{navyblue}{rgb}{0.0, 0.0, 0.5}
\usepackage{latexsym}
\usepackage{booktabs}
\usepackage[protrusion=true, expansion=true]{microtype}
\usepackage[colorlinks=true, breaklinks=false, linkcolor=navyblue, urlcolor=navyblue, citecolor=navyblue]{hyperref}
\usepackage{float}
\usepackage{mathtools}
\usepackage{braket}
\usepackage{siunitx}
\usepackage{appendix}
\makeatother

\begin{abstract}
     A temperature gradient $\Delta T$ across a Josephson junction induces a thermoelectric current. We predict the AC Josephson effect is activated when this current surpasses the junction's critical current. Our investigation of this phenomenon employs the time-dependent Ginzburg–Landau theory framework in proximity to the critical temperature. Our results indicate that the frequency of the AC current is approximately given by $\pi S \Delta T / (2 \Phi_0)$,  where  $S$ represents the Seebeck coefficient and $\Phi_0$ the magnetic flux quantum and  we estimate the frequency be on the range of GHz for Sn up to a THz for larger $S$ and $T_c$ materials. Furthermore, we propose two distinct experimental configurations to observe this effect.
\end{abstract}

\begin{document}

\title{Thermoelectric AC Josephson effect}
\author{Olli Mansikkamäki}
\affiliation{Nordita,
Stockholm University and KTH Royal Institute of Technology,
Hannes Alfvéns väg 12, SE-106 91 Stockholm, Sweden}
\author{Francesco Giazotto}
\affiliation{NEST, Istituto Nanoscienze-CNR \& Scuola Normale Superiore,
Piazza San Silvestro 12, I-56127  Pisa, Italy}
\author{Alexander Balatsky}
\affiliation{Nordita,
Stockholm University and KTH Royal Institute of Technology,
Hannes Alfvéns väg 12, SE-106 91 Stockholm, Sweden}
\affiliation{Department of Physics, University of Connecticut, Storrs, Connecticut 06269, USA}

\maketitle

In superconducting systems, electrical charge is conveyed by a superfluid composed of Cooper pairs alongside quasiparticles, which are excitations analogous to electrons \cite{tinkham_introduction_2004}. These quasiparticles replicate the behavior of holes or electrons, similar to what is observed in a conventional metal. In particular, within the scope of our study, this involves the Seebeck effect \cite{ashcroft1976solid,abrikosov2017fundamentals,mamin1984charge,galperin2002theory,ginzburg1978thermoelectric,shelly2016resolving,marchegiani2020nonlinear}. A temperature gradient $\Delta T = (T_2 - T_1)$ across a superconducting wire induces a quasiparticle thermoelectric current $j_q \sim \sigma S \nabla T$, where $\sigma$ refers to the normal-state conductivity and $S$ denotes the Seebeck coefficient. 
The current induced by a temperature gradient exhibits a formal similarity to the current elicited by a voltage bias. Therefore, we predict the thermoelectric AC Josephson effect, where a Josephson junction exposed to a temperature gradient would demonstrate behavior analogous to that of a Josephson junction exposed to a voltage bias. Induced current oscillations occur only after a critical value of the thermal difference is reached, which makes it a threshold phenomenon.   We call the predicted effect the {\em thermoelectric AC Josephson effect (TEACJ)}. Electric alternating current (AC) should be observed, with the thermoelectric AC frequency
\begin{equation}\label{eq:phys_freq}
    \Omega_{\text{TE}} = \frac{\pi S \Delta T}{2 \Phi_0}.
\end{equation}

\begin{figure}
    \centering
    \includegraphics[width=\linewidth]{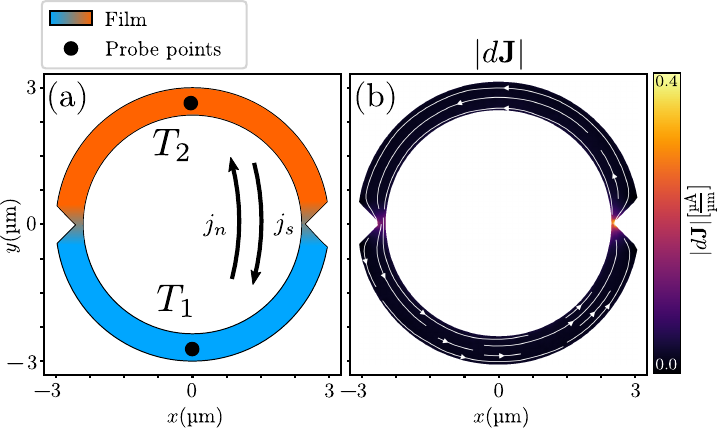}
    \caption{A schematic illustration of a superconducting ring incorporating two different Dayem bridges, constituting a thermally-biased  DC SQUID. The colors depict the temperature distribution within the ring, where $T_1 < T_2 < T_c$ is indicated, with blue representing $T_1$ and orange representing $T_2$. The arrows denote the direction of the thermoelectric quasiparticle current $j_q$ and the compensating dissipationless supercurrent $j_s$. In subsequent figures, phase and voltage differentials are calculated at the probe points indicated by black dots. The total sheet current density $d \mathbf{J}$ is presented in panel (b), with arrows indicating the current's direction.}
    \label{fig:ring}
\end{figure}

\begin{figure}[t!]
    \centering
    \includegraphics[width=\linewidth]{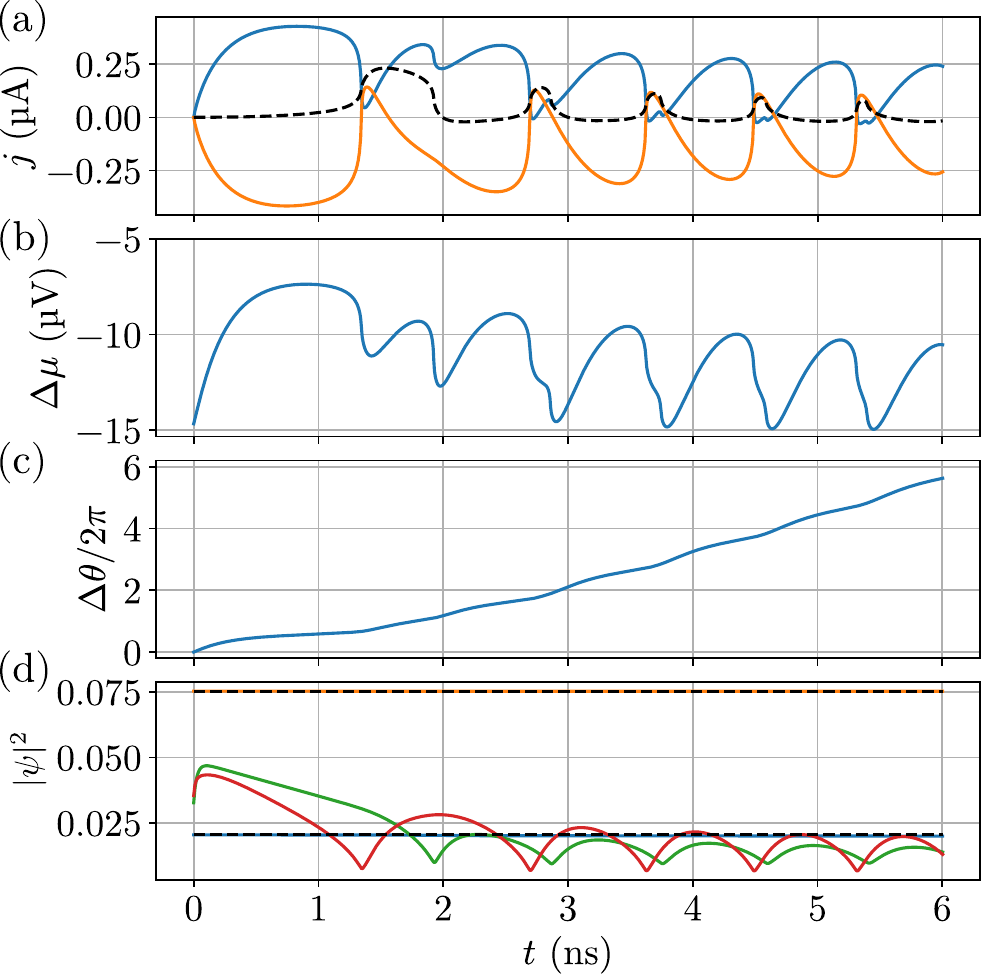}
    \caption{(a) The plots illustrate the quasiparticle current $j_q$ (depicted in solid blue), the supercurrent $j_s$ (depicted in solid yellow), and the aggregate current $j = j_q + j_s$ (depicted in dashed black) traversing the ring. (b) The thermochemical potential difference $\Delta \mu$ is presented between the probe points as detailed in Fig.~\ref{fig:ring}(a). (c) The phase difference $\Delta \theta$ is delineated between the aforementioned probe points. The corresponding temperatures are $T_1 = 0.93 T_c$ and $T_2 = 0.98 T_c$. (d) The superfluid densities $|\psi|^2$ are depicted at the probe points, with yellow representing the bottom and blue the top, alongside the depiction of the right Dayem bridge in red and the left one in green. The black dashed lines denote $\epsilon_1 = T_c / T_1 - 1$  and $\epsilon_2 = T_c / T_2 - 1$.}
    \label{fig:currents}
\end{figure}

\begin{figure}[t!]
    \centering
    \includegraphics[width=\linewidth]{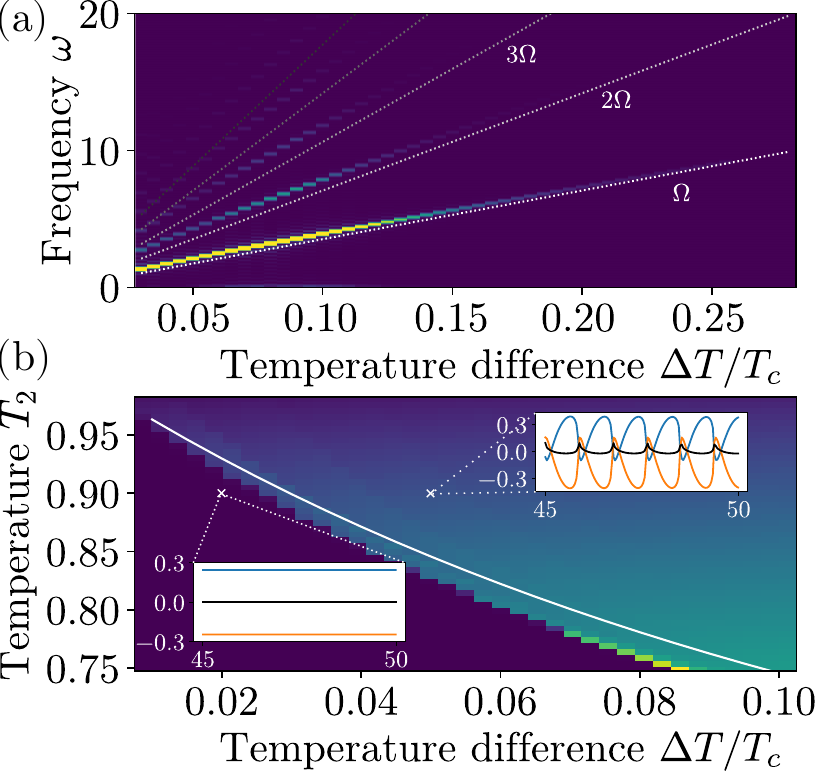}
    \caption{(a) Approximate normalized power spectral density in the form of Lomb--Scargle periodograms~\cite{vanderplas_understanding_2018} for a range of temperature differences $\Delta T / T_c$. The dashed lines depict integer multiples of the frequencies $\Omega$ from Eq.~\eqref{eq:freq}. The first three of the lines are labeled. The higher temperature is $T_2 = 0.98 T_c$. (b) The amplitude $A$ of the total current is presented as a function of the higher temperature $T_2$ and the temperature difference $\Delta T$. The white line indicates the temperatures at which the numerator of Eq.~\eqref{eq:freq} equals zero. The current oscillations occur only above the critical temperature difference. The insets show the currents between $\SI{45}{\nano\second}$ and $\SI{50}{\nano\second}$ at the temperature $T_2 = 0.9 T_c$ and the differences $\Delta T = 0.02 T_c$ and $\Delta T = 0.05 T_c$, marked with white crosses. A video showing the time evolution of the currents until $\SI{50}{\nano\second}$ is available in the supplemental material~\cite{supplement}. (temporary link: \href{https://stockholmuniversity.box.com/s/0mhojh3lhu9s4tj5s4l31jko07d657e1}{link})}
    \label{fig:steady}
\end{figure}

To illustrate the effect, we investigate the dynamics of a thermally biased superconducting ring incorporating two Dayem bridges (refer to Fig.~\ref{fig:ring}), utilizing the time-dependent Ginzburg-Landau theory~\cite{tinkham_introduction_2004}. The total resulting oscillating AC current for a given $\Delta T$ is shown in Fig.~\ref{fig:currents}. Equation~\ref{eq:phys_freq} along with the spectral analysis of current oscillations in Figs.~\ref{fig:currents} and~\ref{fig:steady} are the main results of this article. 

Before proceeding to the detailed calculations, we underscore several important points: {\bf i}) The effect we predict is qualitatively distinct from the traditional AC Josephson effect. In the conventional scenario, coherent oscillations occur between two condensates with voltage bias $V$, with a condensate that facilitates coherent Rabi-like oscillations between them.
In this scenario, we observe the phenomenon of $eV \rightarrow \hbar \omega_J$ conversion (that is, $\hbar \omega_J = 2e V$), where a coherent voltage shift is applied, resulting in the generation of coherent oscillations.
In the former thermally bias scenario, coherent current oscillations occur while the upper and lower arms are maintained at distinct temperatures. Thus, in the context of a TEACJ, the sole non-equilibrium driver comprises two thermal distributions at temperatures $T_1$ and $T_2$. These distributions are characterized as purely thermal steady-state distributions with inherent thermal fluctuations. However, the superconducting state manages to produce a coherent response at a {\em single frequency} $\Omega_{TE}$. In this way, incoherent thermal distributions give rise to coherent oscillations $(T_1 , T_2) \rightarrow \hbar \Omega_{TE}$ along with harmonics. The Josephson junction therefore reduces the entropy of the input to yield a sharply defined low-entropy output. We identify this result as a significant attribute of the Josephson junction.
{\bf ii}) The anticipated results for TEACJ are expected to be applicable to tunnel junctions as well. Due to limitations inherent in the time-dependent Ginzburg--Landau model and its computational implementation, we employ Dayem bridges.
{\bf iii}) The time--dependent Ginzburg--Landau theory is fundamentally applicable only near the critical temperature $T_c$. Although it can be derived exclusively from microscopic theory in the limit $T \approx  T_c$, empirical evidence indicates that it yields results consistent with experimental observations over a wide range of temperatures~\cite{liu_kinetics_1991, kato_computer_1991, frahm_flux_1991, machida_direct_1993}. We expect the effect to be general and persist in the whole temperature range. The calculations away from the critical temperature would have to be done using microscopic calculations and will be presented elsewhere. 

We modeled the device using the generalized time-dependent Ginzburg--Landau theory~\cite{kramer_theory_1978, watts-tobin_nonequilibrium_1981} as implemented in the open-source py-TDGL package~\cite{horn_pytdgl_2023}, which we modify slightly to include thermoelectric currents. The time-evolution of the complex order parameter $\psi(\mathbf{r}, t) = |\psi|e^{i \theta}$ is governed by the equation
\begin{equation}\label{eq:tdgl}
\begin{split}
     \frac{u}{\sqrt{1 + \gamma^2 |\psi|^2}} &\left( \frac{\partial}{\partial t} + i \mu + \frac{\gamma^2}{2} \frac{\partial |\psi|^2}{\partial t} \right) \psi \\
    &= (\epsilon - |\psi|^2) \psi + (\nabla - i \mathbf{A})^2 \psi.   
\end{split}
\end{equation}
Here, $u = \pi^4 / [14 \zeta(3)]$, where $\zeta$ is the Riemann zeta function, is the ratio of the relaxation times of the amplitude and phase of the order parameter $\psi$ in dirty superconductors. The effect of inelastic electron-phonon scattering is included via the parameter $\gamma = 2 \tau_E \Delta_0$, where $\tau_E$ is the inelastic scattering time and $\Delta_0$ is the zero-field superconducting gap. The effects of electromagnetic fields are given by the electrochemical scalar potential $\mu$, and the magnetic vector potential $\mathbf{A}$. Local variations in temperature are set via the parameter $\epsilon(\mathbf{r}) = T_c / T(\mathbf{r}) - 1$. 

The supercurrent density is given by 
\begin{equation}\label{eq:js}
    \mathbf{J}_s = \text{Im}[\psi^* (\nabla - i \mathbf{A}) \psi], 
\end{equation}
and the quasiparticle current density by
\begin{equation}\label{eq:jq}
    \mathbf{J}_q = -\nabla \mu - \frac{\partial \mathbf{A}}{\partial t} + \eta \nabla \left( \frac{T}{T_c} \right),
\end{equation}
where $\eta$ is a dimensionless parameter given by 
\begin{equation}
    \eta = \frac{\pi \mu_0 \lambda^2 \sigma S T_c}{2 \Phi_0},
\end{equation}
where $\lambda$ is the magnetic penetration depth, $\sigma$ is the normal state conductivity, $S$ is the Seebeck coefficient, $T_c$ is the superconducting critical temperature, $\mu_0$ is the vacuum permeability, and $\Phi_0$ is the magnetic flux quantum. The last term of Eq.~\eqref{eq:jq} containing the thermoelectric contribution $\eta \nabla \left( \frac{T}{T_c} \right)$ to the current density is a deviation from the implementation of Ref.~\cite{horn_pytdgl_2023}. We assume that there is no external electric field and that the charge density is approximately locally conserved~\cite{watts-tobin_nonequilibrium_1981}, that is, the continuity equation $\nabla \cdot (\mathbf{J}_s + \mathbf{J}_q) = 0$ holds. From this assumption and the above expressions for the current densities, we can derive a Poisson equation 
\begin{equation}\label{eq:poisson}
    \nabla^2 \mu = \nabla \cdot \mathbf{J}_s - \nabla \cdot \frac{\partial \mathbf{A}}{\partial t} + \eta \nabla^2  \left( \frac{T}{T_c} \right),
\end{equation}
from which we compute the electrochemical potential. 

In the following, we assume the vector potential $\mathbf{A} = 0$ unless otherwise specified. We simulate a superconducting ring, shown in Fig.~\ref{fig:ring} with the inner radius $\SI{2.4}{\micro\meter}$ and the outer radius $\SI{3}{\micro\meter}$. The ring has two Dayem bridges~\cite{dayem_behavior_1967} with minimum widths $\SI{90}{\nano\meter}$ and $\SI{30}{\nano\meter}$. Note that the SQUID needs to be \textit{asymmetric} to support a non-zero total current circulating in the ring. We set the upper half to the temperature $T_2 / T_c = 0.98$ and vary the temperature $T_1$ of the lower half. 
In the experimental configuration, it is recommended to replace the current junctions with SIS tunnel junctions to reduce thermal conduction between the two segments of the SQUID.

While Eq.~\eqref{eq:tdgl} is presented here in a dimensionless form, the time and length scales of the dynamics are given by the properties of the material of the modeled device. In our simulations, we use parameters similar to those measured in thin films of tin (Sn), as it has been estimated to have a relatively high $\sigma S =\SI{54}{\volt /(\kelvin \cdot\ohm \cdot\centi\meter)}$~\cite{zavaritskii_observation_1974}. 
In principle, all the parameters are temperature-dependent. Still, for the sake of simplicity and due to the limitations of the numerical implementation, we take the parameters to be uniform throughout the ring. We take the film to have the thickness $d = \SI{40}{\nano\meter}$. For such a film, we get the effective penetration depth $\lambda(T_2, d) \approx \SI{465}{\nano\meter}$~\cite{douglass_precise_1962} and the coherence length $\xi(T_2) \approx \SI{365}{\nano\meter}$~\cite{douglass_precise_1962}. Using the value $T_c = \SI{3.88}{\kelvin}$ for the critical temperature of Sn~\cite{dolan_critical_1974}, we get $\eta \approx 4.33$, with $\rho = 1 / \sigma = \SI{1.4}{\micro\ohm \cdot \centi\meter}$~\cite{douglass_precise_1962}. Assuming the gap scales as $\frac{\Delta_0(T)}{\Delta_0(0)} \approx 1.74 \left(1 - \frac{T}{T_c}\right)^{\frac{1}{2}}$~\cite{tinkham_introduction_2004}, we set $\gamma = 97$, taking the inelastic scattering rate as $\tau_E = \SI{0.2}{\nano\second}$~\cite{hsiang_boundary_1980} and the gap $\Delta_0 \approx \SI{1}{\milli\electronvolt}$\cite{townsend_investigation_1962}.

We use the initial state $\psi = \sqrt{\epsilon}$ and keep the temperatures constant in our simulations. The temperature is effectively ``switched on'' instantaneously, resulting in a sharp jump in currents at the beginning of simulations. The initial behavior of the currents, voltage, phase difference and superfluid density $|\psi|^2$ is shown in Fig.~\ref{fig:currents}. To understand the dynamics, let us first examine how the temperature difference affects the currents. Most obviously, in Eq.~\eqref{eq:jq}, we have the direct thermoelectric contribution to the quasiparticle current density
\begin{equation}
    \mathbf{J}_{q, T} = \eta \nabla \left( \frac{T}{T_c} \right),
\end{equation}
as a result of the temperature gradient, also known as the Seebeck effect. 

The temperature gradient also affects the density of the superfluid $|\psi|^2$ through the $(\epsilon - |\psi|^2) \psi$ term of Eq.~\eqref{eq:tdgl}. This means that the superfluid density tends to relax towards the distribution $|\psi|^2 \rightarrow \epsilon = \frac{T_c}{T} - 1$. A non-uniform temperature distribution results in a non-uniform superfluid density, which directly affects the supercurrent density, which in the absence of an external magnetic field is given by
\begin{equation}
    \mathbf{J}_s = |\psi|^2 \nabla \theta.    
\end{equation}
The higher the temperature, the less superfluid there is to contribute to the supercurrent. The Poisson equation~\eqref{eq:poisson} for the thermochemical potential $\mu$ also contains a term describing the temperature distribution. Since the time evolution of the phase $\theta$ of the order parameter, as governed by Eq.~\eqref{eq:tdgl}, depends on the thermochemical potential, a temperature difference across the Dayem bridges implies that there must also be a phase difference.

When discussing the AC Josephson effect, it is common to simplify the current-phase relation as
\begin{equation}\label{eq:josephson}
    j(\Delta\theta) = j_c \sin(\Delta \theta).
\end{equation}
Generally, the oscillation is not sinusoidal but features higher harmonics $\sin(2 \Delta \theta)$, $\sin(3 \Delta \theta)$, and so on~\cite{golubov_current-phase_2004, goldobin_josephson_2007, de_lange_realization_2015, stoutimore_second-harmonic_2018, kringhoj_anharmonicity_2018, willsch_observation_2024}. The higher harmonics are also apparent in our system, as shown in Fig.~\ref{fig:steady}(a). 
However, for simplicity, we assume that Eq.~\eqref{eq:josephson} describes the oscillation corresponding to the highest peak. 

Let us look at the oscillation frequency of the AC current by greatly simplifying the model. To this end, we consider the system only at the probe points shown in Fig.~\ref{fig:ring}, far from the junctions.  We assume that the superfluid around the points is uniformly distributed and $\nabla^2 |\psi_k| = 0$. After the initial transient behavior, the magnitude of the order parameter at the probe point $k$ is constant $\frac{\partial |\psi_k|}{\partial t} = 0$. Then, the real part of Eq.~\eqref{eq:tdgl} is simply
\begin{equation}
    (\epsilon - |\psi|^2) |\psi| = 0,
\end{equation} 
and we get the superfluid densities $|\psi_k|^2 = \epsilon_k$. We can see in Fig.~\ref{fig:currents} (d) that this is approximately correct. From the imaginary part of Eq.~\eqref{eq:tdgl} we get the time evolution of the phase difference between the points
\begin{equation}\label{eq:phase_evolution}
    \frac{\partial \Delta \theta}{\partial t} = -\Delta\mu.
\end{equation}
To estimate the thermochemical potential difference $\Delta \mu$, we use a discretization scheme similar to the numerical implementation of Ref.~\cite{horn_pytdgl_2023}, in which
\begin{equation}\label{eq:potential}
\begin{split}
    \Delta \mu &= \eta \frac{\Delta T}{T_c} + \text{Im} [|\psi_1| e^{-i \theta_1} (|\psi_2| e^{i \theta_2} - |\psi_| e^{i \theta_1})]  \\
    &= \eta \frac{\Delta T}{T_c} + |\psi_1| |\psi_2| \sin(\Delta \theta).
\end{split}
\end{equation}
Combining Eqs.~\eqref{eq:phase_evolution} and \eqref{eq:potential}, we get a somewhat complicated expression for the phase difference,
\begin{equation}\label{eq:phase_difference}
    \Delta \theta = 2 \tan^{-1} \left[ \frac{\alpha \tan \left(\frac{t}{2} \alpha + 2 \tan^{-1} \left( \frac{\beta}{\alpha}\right) \right) - \beta}{\eta \frac{\Delta T}{T_c}}\right],
\end{equation}
where $\alpha = \sqrt{\eta^2 (\frac{\Delta T}{T_c})^2 - \epsilon_1 \epsilon_2}$, and $\beta = \sqrt{\epsilon_1 \epsilon_2}$. If we ignore the time-independent terms, we can estimate the oscillation frequency of the AC current as
\begin{equation}\label{eq:freq}
    \Omega = \frac{\eta^2  (\frac{\Delta T}{T_c})^2 - \epsilon_1 \epsilon_2}{\eta  \frac{\Delta T}{T_c}}.
\end{equation}
Further, in the limit $\eta^2  (\frac{\Delta T}{T_c})^2 \gg \epsilon_1 \epsilon_2$, which holds when the higher temperature approaches the critical temperature, we can ignore the $\epsilon_1 \epsilon_2$ term and estimate the thermoelectric AC frequency, in physical units, simply as
\begin{equation}
    \Omega_\text{TE}\approx \frac{\eta \frac{\Delta T}{T_c}}{\tau_0} = \frac{\pi S \Delta T}{2 \Phi_0},
\end{equation}
where $\Omega_\text{TE} = \Omega / \tau_0$, and $\tau_0 = \mu_0 \sigma \lambda^2$ is the characteristic timescale of the time-dependent Ginzburg--Landau model. The approximate frequency $\Omega$ is compared to the numerical solution in Fig.~\ref{fig:steady}(a).

We estimate the TEACJ frequency to be in the range of $\Omega_{TE} \sim 1-10 Ghz$  for  characteristic parameters of Sn thin films with Seebeck coefficient $S \sim 75  {\mu V K^{-1}} $. The oscillation frequency and the radiation emitted are well aligned with the experimental setups mentioned below. We also estimate the frequency for FeSe.  FeSe superconducting films demonstrate substantially higher critical temperatures $T_c = \SI{65}{\kelvin}$ \cite{wang2012interface, liu_electronic_2012}, and potentially even in the range of $\SI{100}{\kelvin}$~\cite{ge_superconductivity_2015}. In addition, the Seebeck coefficient of these films has been recorded to reach values as high as $S = \SI{454}{\micro\volt\kelvin^{-1}}$~\cite{shimizu_giant_2019} above 50 K. The estimated frequency is then in the range of $\Omega_{TE} \sim 1 Thz$, thus making FeSe a promising candidate for thermoelectric applications within superconductors.

Note that the solution shown in Eq.~\eqref{eq:phase_difference} is valid as a phase difference only when $\alpha$ has a real value. That is, we do not expect to see an AC current when the numerator of Eq.~\eqref{eq:freq} is negative. To compare this prediction to numerical results, we plot
\begin{equation}\label{eq:amplitude}
    A = \frac{1}{2}(j_{\text{max}} - j_{\text{min}}),
\end{equation}
where $j_{\text{max}}$ and $j_{\text{min}}$ are the maximum and minimum values of the total current, as a function of the higher temperature $T_2$ and the temperature difference $\Delta T$, as shown in Fig.~\ref{fig:steady}(b). For an oscillating solution, $A$ is the amplitude of the oscillation of the total current, and for a steady state, $A = 0$.

Although the simple approximation matches the numerical results quite well, it should be noted that we have assumed the material parameters to be constant throughout the ring. In reality, they would depend quite strongly on the temperature; consequently, the frequency $\Omega$ of the alternating current is improbable to be quantitatively precise, except in proximity to the critical temperature.

\begin{figure}
    \centering
    \includegraphics[width=\linewidth]{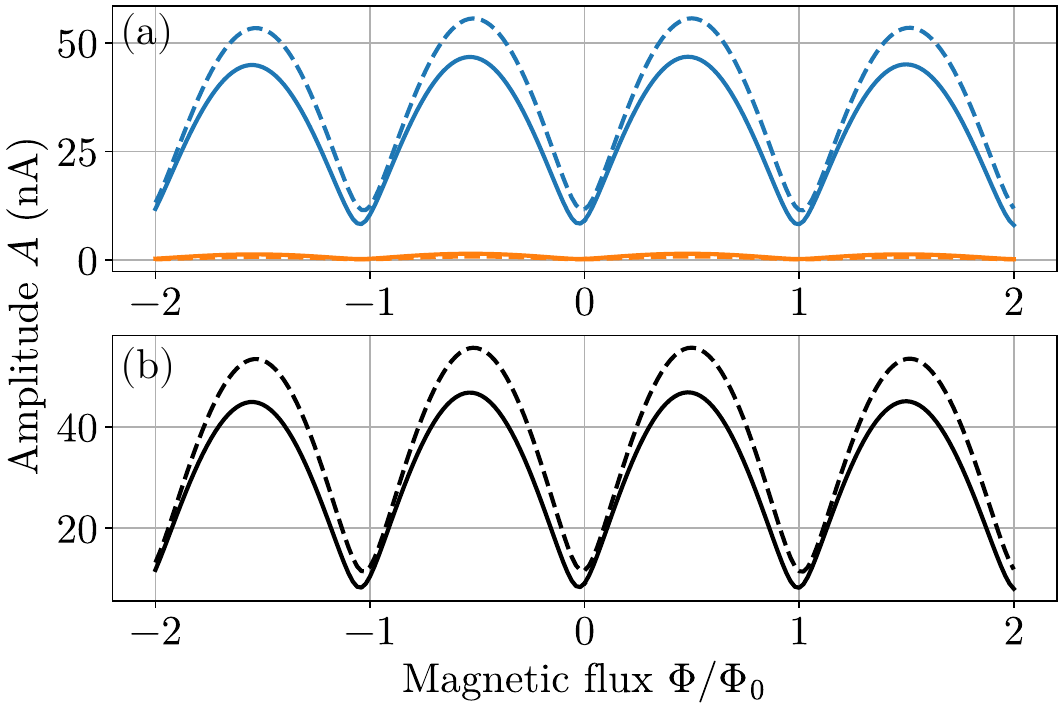}
    \caption{The amplitude of alternating current (AC) through the ring is analyzed as a function of the magnetic flux $\Phi / \Phi_0$ through the ring. The amplitude $A$ is evaluated following the definition provided in Eq.~\eqref{eq:amplitude} throughout $\SI{1}{\nano\second}$ after the alternating current has stabilized. Panel (a) showcases the supercurrent $j_s$ depicted in yellow and the quasiparticle current $j_q$ displayed in blue. Panel (b) represents the total current $j = j_s + j_q$. The higher temperature is set at $T_2 = 0.98 T_c$. In both panels, the case $\Delta T = 0.1$ is shown in solid and $\Delta T = 0.2$ in dashed lines. The slight shift to the left along the horizontal axis is caused to the non-zero flux due to the thermoelectric current. Flipping the temperatures $T_1$ and $T_2$ results in a shift to the right. }
    \label{fig:field}
\end{figure}

Let us now drop the assumption $\mathbf{A} = 0$ and see how an applied magnetic field affects the current. We assume that the magnetic field is constant, $\frac{\partial \mathbf{A}}{\partial t} = 0$. Since a constant field does not contribute to the potential $\mu$ via Eq.~\eqref{eq:poisson}, it does not affect the frequency of the AC current. However, it affects the amplitude of the oscillating current, as shown in Fig.~\ref{fig:field}. 
The total flux through a ring with radius $r = \SI{2.7}{\micro\meter}$ is $\Phi_0$ when the magnetic field strength is $\mu_0 H = \frac{\Phi_0}{\pi r^2} = \SI{90.25}{\micro\tesla}$. We set a uniform magnetic field $\mu_0 H$ along the z axis and vary its strength from $\SI{-180.5}{\micro\tesla}$ up to $\SI{180.5}{\micro\tesla}$. We observe a significant change in the amplitude of the total current, with a maximum near $\mu_0 H = \SI{45.14}{\micro\tesla}$, at a total flux of $\Phi_0 / 2$. 

\begin{figure}
\centering
\includegraphics[width=\linewidth]{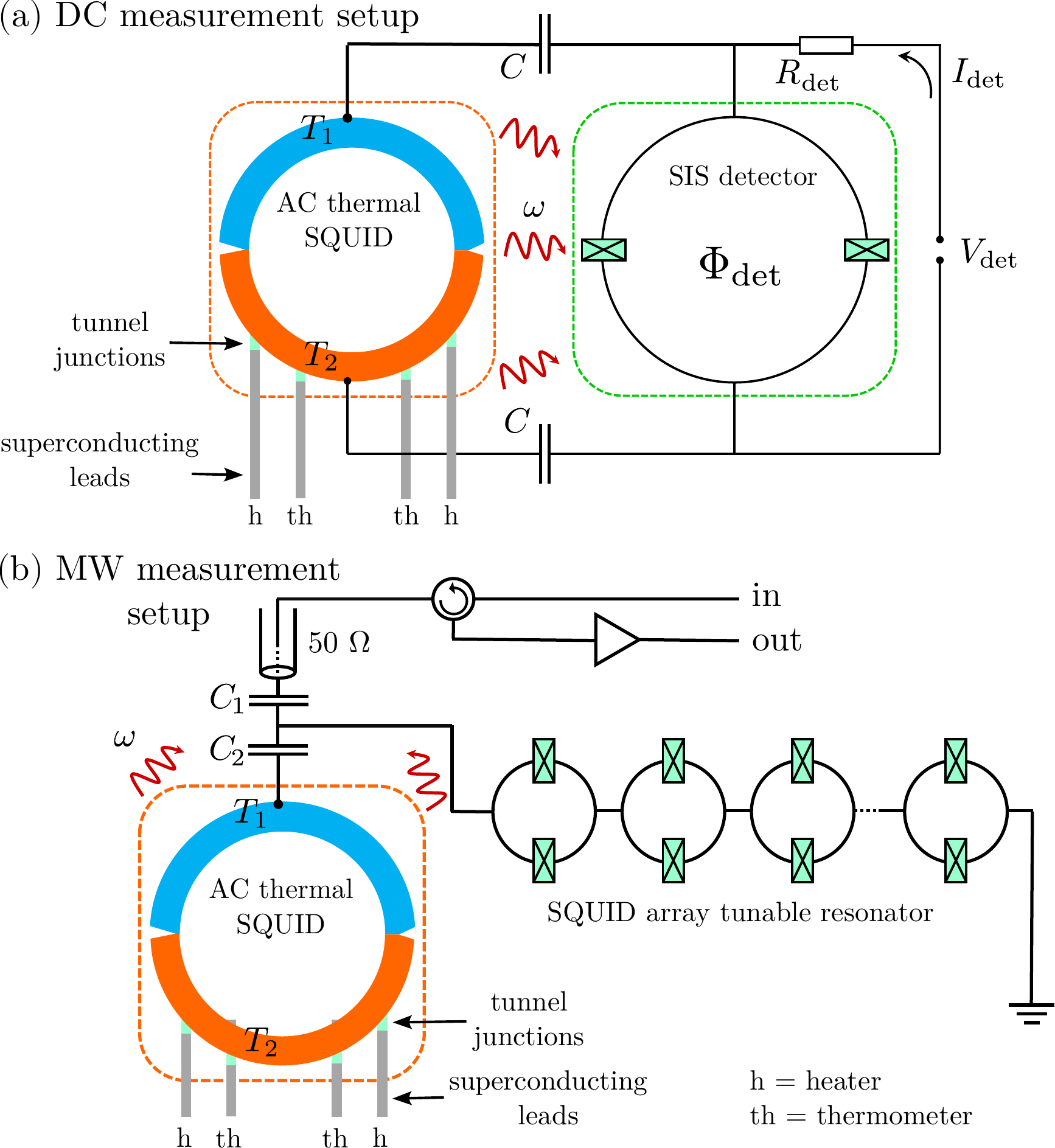}
        
\caption{Schemes of possible measurement setups for the TEACJ effect. (a) A DC measurement setup features a thermally-biased superconducting quantum interference device (SQUID) (orange box). Additional superconducting leads, tunnel-coupled to the bottom half of the interferometer, enable the imposition and measurement of the temperature gradient across the superconducting loop. These junctions function as heaters (h) and thermometers (th). The electromagnetic radiation at frequency $\omega$ arising from the TEACJ effect is detected by capacitively coupling the left SQUID to a superconductor-insulator-superconductor (SIS) DC SQUID on the right (green box), which serves as an on-chip radiation detector. The total Josephson coupling in the SIS SQUID can be minimized using the external magnetic flux $\Phi_{det}$ to investigate the photon-assisted quasiparticle tunneling current contribution to the SQUID DC characteristic. $V_{det}$ is the applied DC bias voltage, with an effective resistance $R_{det}$ that results in a current $I_{det}$. $C$ denotes the coupling capacitors.
(b) A microwave (MW) measurement setup consists of a thermally-biased SQUID coupled to a frequency-tunable superconducting resonator realized with an array of SIS SQUIDs. $C_1$ and $C_2$ represent the coupling capacitance to the MW detection chain and the resonator, respectively. The MW detection chain, which includes essential components like the circulator and preamplifier, performs reflectance measurements through conventional circuit quantum electrodynamics heterodyne detection techniques.}
\label{fig:exp}
\end{figure}

In contrast to the simulations presented here, an experimental setup will be subject to thermal current noise, potentially rendering the oscillations undetectable. Assuming that the thermal noise approximates the magnitude $\tilde j = \frac{2 e k_B T}{h}$, the temperature of the upper half of our simulated Sn ring is $T_2 = 0.98 T_c = \SI{3.802}{\kelvin}$, which results in a thermal noise magnitude of $\tilde j \approx \SI{0.025}{\micro\ampere}$. As illustrated in Fig.~\ref{fig:field}, without an external magnetic field and the lowest half temperature at $T_1 = 0.88 T_c$, the noise is comparable to the amplitude of the total current, thus hindering detection. However, as demonstrated, the amplitude can be considerably enhanced by either reducing the temperature $T_1$ [refer to Fig.~\ref{fig:steady} (b)] or applying an external magnetic field [see Fig.~\ref{fig:field}], while neither method is expected to increase the thermal noise. Therefore, we contend that thermal noise should not impede the experimental detection of the AC current.

Figure \ref{fig:exp} shows two possible experimental configurations devised to evaluate the TEACJ. These configurations can be constructed using conventional lithography methods. The arrangement in (a) illustrates a DC measurement approach wherein the thermally-biased SQUID is capacitively coupled to a superconductor-insulator-superconductor (SIS) DC SQUID, functioning as an on-chip detector for the radiation emitted at frequency $\omega$. A loop segment is joined to four additional superconducting electrodes via oxide barriers. These electrodes heat the quasiparticles in the superconductor and act as sensitive thermometers to accurately gauge the thermal gradient imposed across the structure \cite{giazotto_opportunities_2006,germanese2022bipolar,giazotto2012josephson,fornieri2017towards}. The overall Josephson coupling within the SIS detector can be reduced through the application of external magnetic flux ($\Phi_{det}$), permitting exploration of the contribution of photon-assisted quasiparticle tunneling current to the DC SQUID current versus voltage characteristic~\cite{deblock_detection_2003, van_woerkom_josephson_2017, laroche_observation_2019}.

The scheme illustrated in Fig. \ref{fig:exp}(b) represents a potential alternative setup based on a microwave (MW) detection architecture, where the thermally-biased SQUID is capacitively connected to a frequency-tunable superconducting resonator made up of an SQUID array \cite{stockklauser_strong_2017}. The resonance frequency can be adjusted using a small magnetic field applied through an on-chip flux line. This MW set-up enables the characterization of emitted radiation through reflectance measurements using standard circuit quantum electrodynamics detection methods.

In summary, within the framework of the time-dependent Ginzburg--Landau theory, we have demonstrated that a thermal gradient between the two halves of a SQUID can induce an alternating current across the junction, termed the thermoelectric AC Josephson (TEACJ) effect. Arguably the most notable observation arising from this assertion is the coherent sharp-line (low-entropy) output produced by the SQUID when incoherent thermal distributions are applied to the arms of the interferometer. This emitted sharp line effectively represents the filtering action of the SQUID on the thermal gradient.
An analytical expression, derived in Eq.~\eqref{eq:freq} for $\Omega_{TE}$, not only provides an approximation of the oscillatory frequency, but also serves as a guide for selecting appropriate temperatures of the SQUID halves for a given thermoelectric coefficient $\eta$. Although based on a simplified two-point model, this expression aligns with numerical simulations across a broad temperature range. 
Considering that the oscillation arises from a dissipative state, the phenomenon is anticipated to be detected via emitted radiation. Furthermore, we propose two experimental configurations to empirically validate our predictions.

The authors thank A. Crippa for the valuable discussions. The work of A.B. and O.M. was supported by the European Research Council under the European Union Seventh Framework ERS-2018-SYG 810451 HERO and by the Knut and Alice Wallenberg Foundation Grant No. KAW 2019.0068.
F.G. acknowledges the EU’s Horizon 2020 Research and
Innovation Framework Programme under Grants No. 964398 (SUPERGATE), No. 101057977 (SPECTRUM),
and the PNRR MUR project PE0000023-NQSTI for partial financial support.

\bibliography{references.bib}
\appendix
\section{Approximate power spectral densities}
\begin{figure}[h]
    \centering
    \includegraphics[width=\linewidth]{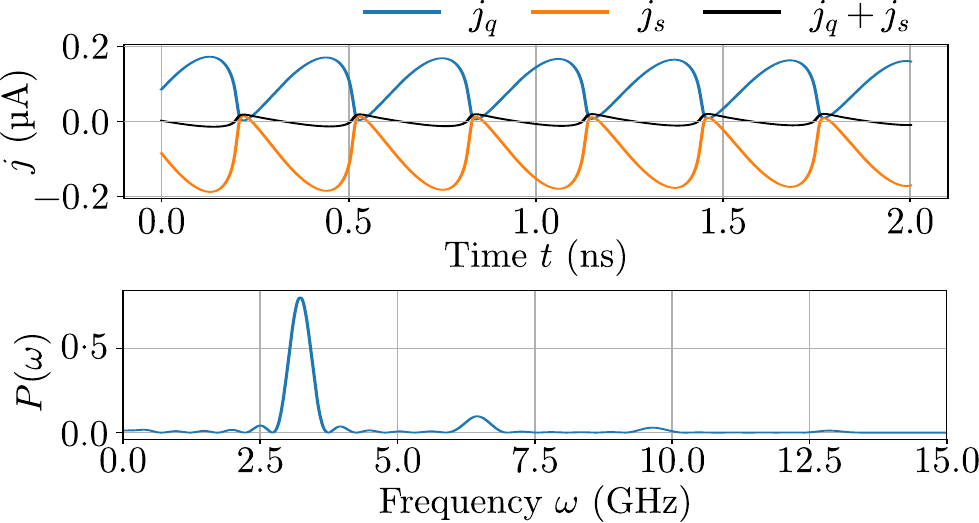}
    \caption{(a) The quasiparticle current $j_q$ (blue), the supercurrent $j_s$ (yellow), and the total current $j = j_q + j_s$ (black) traversing the ring. (b) The Lomb-Scargle periodogram $P(\omega)$ correspondinc to the total current shown in (a). The temperature difference is $\Delta T = 0.2 T_c$ and the first $\SI{10}{\nano\second}$ are omitted.}
    \label{fig:lombscargle}
\end{figure}

In Fig.~\ref{fig:steady}(a), Lomb--Scargle periodograms~\cite{vanderplas_understanding_2018} are used to display the dependence of the spectral density on the temperature difference. The periodograms give a good estimate for the spectral densities with a relatively small number of unevenly spaced data points, allowing us to speed up the simulations using an adaptive time-step. After the initial behavior, shown in Fig.~\ref{fig:currents} (a). The frequency of the alternating current settles quickly, but the amplitude does not due to the large value of $\gamma$. Since we are mainly interested in the frequency of the highest peak of the spectral density, this is not an issue. An example of the evened out alternating current and the corresponding periodogram is shown in Fig.~\ref{fig:lombscargle}.

\end{document}